# Credit Portfolio Management in a Turning Rates Environment

**Arthur M. Berd**
General Quantitative

**Elena Ranguelova**
Investcorp

**Antonio Baldaque da Silva**
Barclays Capital

*We give a detailed account of correlations between credit sector/quality and treasury curve factors, using the robust framework of the Barclays POINT® Global Risk Model. Consistent with earlier studies, we find a strong negative correlation between sector spreads and rate shifts. However, we also observe that the correlations between spreads and Treasury twists reversed recently, which is likely attributable to the Fed's ongoing quantitative easing. We also find that short-term effective durations in the banking industry are now significantly lower than historical patterns would indicate. Our findings are relevant for credit portfolio managers contemplating the impact of rising interest rates and steepening Treasury curve on corporate bond portfolios.*

## 1. INTRODUCTION

The gradual recovery of the U.S. economy from the consequences of the financial crisis has brought the prospect of the Fed ending its extraordinary quantitative easing (QE) policies. The "moderate tapering" of the rate of QE, pre-announced in May 2013, has jolted the bond market and perhaps marked the turning point in the interest rates from historically low levels. Although the timing might still be uncertain, the expected eventual rise in rates has come to the forefront of many investors' concerns.

Over the past few years, we have witnessed an (albeit slow) economic recovery and a concurrent emergence of a benign credit cycle associated with tight spreads and low volatility. The management of credit portfolios in such an environment requires a more precise positioning with respect to the movements of the underlying interest rates, as the credit-specific spread movements become less pronounced and the impact of systemic factors becomes relatively more important.

It is a widely held belief among credit bond portfolio managers that rates and spreads are negatively correlated. The main fundamental reason is that both US Treasury yields and credit spreads reflect the state of the economy, and therefore one can expect their changes to be correlated to the extent that they are caused by the same underlying economic expectation. A worsening economy is generally associated with falling rates, while an improving economy is associated with rising overall level of interest rates. For spreads the direction of the dependence is precisely the opposite – spreads rise when the economy deteriorates and default risk rises, and they tighten as the economic conditions improve. Accordingly, analysts find negative correlation between corporate bond spreads and US Treasury yields (see Ng, Phelps and Lazanas (2013) for a recent look into this issue).

The above statement on negative correlation applies only to overall changes in Treasury rates, i.e. to "parallel shifts" of the Treasury curve. However, the shape of the yield curve can change in a much more complex way, including twists and butterflies. The dependence of spreads on such changes in the underlying yield curve is much less documented. In terms of economic as well as statistical significance, the parallel shifts and (flattening or steepening) twists are the primary modes of change of the Treasury curve, explaining more than 80% of its variability. Therefore, we focus on these factors and their impact on credit spreads.

In this article we revisit the analysis of the co-movement between the interest rates and spreads originally published in 2003-2004 (see Berd and Ranguelova (2003) and Berd and Silva (2004)). We analyze the relationship between US interest rates and credit spreads using the statistically robust framework of the Barclays POINT® Global Risk Model (see Lazanas et al. (2011)).

We confirm the strong evidence that rates and spreads are negatively correlated: higher rates are associated with tighter spreads and steeper credit curves while lower rates are associated with wider spreads and flatter credit curves across all industries. The change in the slope of the treasury yield curve has a different effect on credit OAS: yield curve flattening typically coincides with narrowing and steepening of credit spread curves, with yield curve steepening having the opposite effect. Furthermore, we observe characteristic differences in the impact of rates on various sectors and on spread curve shapes and OAS dispersion.

Our results are qualitatively robust to different periods of analysis and different data calibration methodologies. However, our findings are conditional on the historical relationship between interest rates and spreads. Managers forecasting a reversal on this stable historical pattern (eg due to QE policies and intervention or increased sovereign risk) will find this analysis less useful[1].

Our findings have significant implications for credit portfolio managers. The negative correlation of spreads with rates affects the duration management of credit portfolios, particularly when there is a significant under- or overweight position with respect to a benchmark containing Treasury bonds. The differential impact across industries and ratings gives rise to potential curve-driven cross-sector relative value opportunities. Finally, the knowledge of spread curve shape and OAS dispersion dependence allows one to derive intra-sector relative value plays based on the outlook for Treasury curves.

## 2. THE CO-MOVEMENTS OF CREDIT SPREADS AND INTEREST RATES

Before presenting the model results, let us define the relevant components of the interest rate curve and illustrate the historical co-movement of credit spreads and interest rates.

### 2.1 Defining the Treasury Curve Shifts and Twists

We define the Treasury shift factor as a uniform increase in the five key-rate factors included in the Barclays POINT® Global Risk Model[2], corresponding to the 2, 5, 10, 20 and 30 year key rates. The Treasury twist factor is defined as a series of changes in the same key-rate factors that correspond to a steepening rotation around the 10-year maturity. See Table 1, where we used a 10 bp scale for shifts and twists as an illustration.

These definitions differ slightly from the statistically more precise approach known as *principal component analysis*. However, they get us pretty close to the true principal components of rates changes and are easier to visualize and discuss.

**Table 1: Treasury Curve Primary Factors**

| Key Rate Maturity (years) | 2 | 5 | 10 | 20 | 30 |
|---|---|---|---|---|---|
| Treasury Curve Shift (bp) | 10 | 10 | 10 | 10 | 10 |
| Treasury Curve Twist (bp) | -10 | -5 | 0 | 5 | 10 |

To explain in practical terms, assume that the yield curve has undergone an arbitrary change in each of its key rate points, denoted as $\Delta y_i$, where the index $i$ corresponds to maturities 2, 5, 10, 20, and 30 years. We can now the approximation to the yield curve change in terms of the shift and twist factors as a linear combination of a unit parallel shift and a unit steepening twist with yet undetermined coefficients and the residual term containing the portion of the yield curve change that is not captured by shift and twist:

$$
\begin{bmatrix} \Delta y_2 \\ \Delta y_5 \\ \Delta y_{10} \\ \Delta y_{20} \\ \Delta y_{30} \end{bmatrix} = \gamma_{shift} \cdot \begin{bmatrix} 1 \\ 1 \\ 1 \\ 1 \\ 1 \end{bmatrix} + \gamma_{twist} \cdot \begin{bmatrix} -2 \\ -1 \\ 0 \\ 1 \\ 2 \end{bmatrix} + \begin{bmatrix} \varepsilon_2 \\ \varepsilon_5 \\ \varepsilon_{10} \\ \varepsilon_{20} \\ \varepsilon_{30} \end{bmatrix}
$$

---

[1] In this regard Eisenthal-Berkovitz et al. (2013) document positive correlation between treasury and credit bonds for some European distressed countries.
[2] We exclude the 0.5 year key rate factor in order to avoid picking up dependencies on peculiar movements of the short end of the Treasury curve.

If we assume that, by construction, the residual term does not contain either a parallel shift of a steepening/flattening component, we can easily find the factor loadings $\gamma_{shift}$ and $\gamma_{twist}$ of the two primary yield curve components as follows:

$$
\gamma_{shift} = \frac{1}{5} \cdot \left( \Delta y_2 + \Delta y_5 + \Delta y_{10} + \Delta y_{20} + \Delta y_{30} \right)
$$

$$
\gamma_{twist} = \frac{1}{10} \cdot \left( -2 \cdot \Delta y_2 - \Delta y_5 + \Delta y_{20} + 2 \cdot \Delta y_{30} \right)
$$

These formulas justify the representation of the shift and twist factors in Figure 1.

### 2.2 Historical Co-Movement of Credit Spreads and Interest Rates

The past two decades were characterized by large shifts and twists of the Treasury yield curve, with the current levels of interest rates just off the historical lows and stands almost 600 bps lower than in 1990, while the curve steepness being close to its historical highs. At the same time, the Barclays Credit Index OAS has experienced wide swings from the tightest levels in low 50 bps in early 1997 to the widest of over 535 bp in November 2008, returning to moderate levels of 140 bps more recently. The corresponding time series are shown in Figure 1, where we show the cumulative time series of the shift and twist factors, normalized to start from zero at the start of January 1990 (on the left axis), and the OAS (on the right axis).

We can identify several periods in the past fourteen years when rate changes have visibly correlated with credit spreads.

First, the first Gulf War in 1991 resulted in a quick drop in Treasury rates by over 80 bps and a moderate 5 bp steepening of the curve. At the same time, the credit OAS widened by almost 60 bps.

The dramatic rates swings in 1994, ignited by the Mexican peso crisis and followed by MBS market problems in the US, saw the Treasury rates shift up by about 250bps, with a simultaneous flattening of the curve by 45 bps. The OAS over the same period remained range bound, eventually widening by 10 bps, almost all of it during the last four months of the year, when most of the Treasury curve flattening also took place.

Then, in August-October 1998 the Treasury curve shifted 100 bps in a negative direction and twist moved 10 bps in a positive direction after the Russian default and LTCM crisis prompted the Fed to cut the short rates. Spreads moved sharply wider by 40 bps, but then reversed just as the yield curve twist subsided and the rates themselves moved higher in the beginning of 1999.

Next, the interest rate curve inverted (twist became negative) in the latter part of 1999 and beginning of 2000 as the FOMC raised the Fed Funds rate up to 6.50%, pushing the 2 year yield to 6.70% while the Treasury buybacks, budget surpluses and dampened inflation expectations helped to keep the long yields subdued at 6.00%. The credit spreads widened through this period by over 60 bps, apparently anticipating the coming risks in the equity markets which were nearing the end of the Nasdaq bubble.

**Figure 1. Treasury Shift and Twist vs. Credit Index OAS, 1990 – 2013**

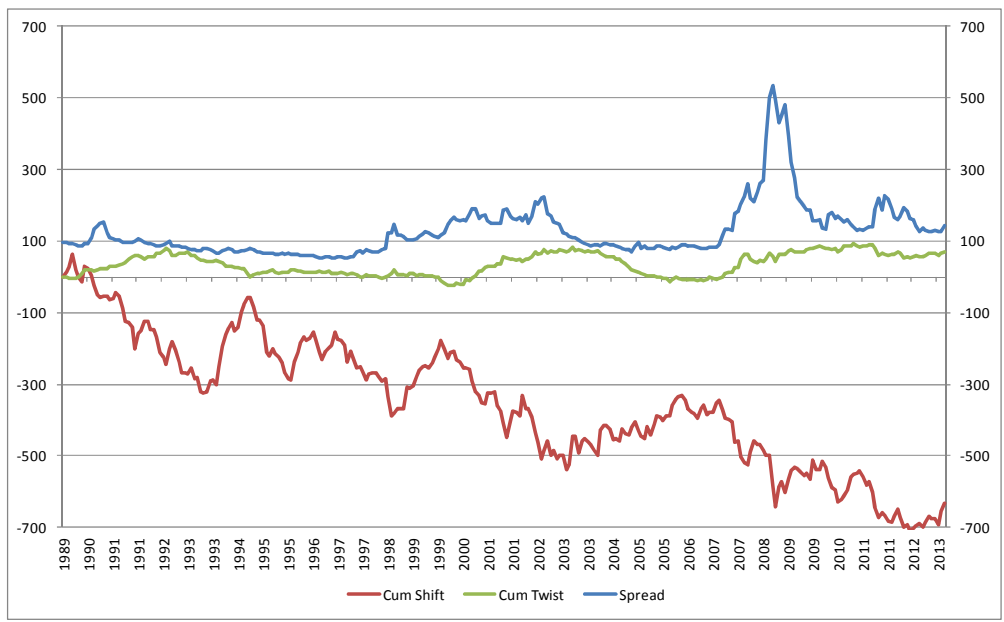

Then the rates curve steadily shifted down (200bp)  and significantly steepened (80bp)from 2000 till 2003 as the Fed cut rates 12 times, from 6.50% to 1.25%. Spreads swung widely for most of this period, from lows close to 100bp to highs above 220bp. The correlation of spreads with the twists becomes significantly positive – in contrast to the negative correlation with the level of rates.. Spreads seemed more sensitive to expectations for the near-term economic outlook, related to the short end of the yield curve, than to long-run growth and inflation, which are related to the long end of the curve.

The economic recovery, accompanied by the rebound in rates (shift higher) and gradual reduction of the steepness of the Treasury curve from 2002 through the beginning of 2006, also saw the credit spreads tightening from over 200 bps to the lowest of 78 bps at the end of this period. Yet again, we see the negative comovement of spreads and Treasury shifts and positive comevement of spreads and twists.

And of course the most dramatic demonstration of this relationship came in 2007-2008 when the unfolding financial crisis pushed the credit spreads to historic highs, while the interest rates were brought further down by both the actions of the Fed and the effects of the economic recession. Between July 2007 and the end of 2008 the curve shifted down by another 300 bps across the board, while steepening by 50 bps.

The emergence from the crisis in 2009 brought a small rebound of the rates up (driven by the long end) but the still ongoing Fed quantitative easing programme has led to further shift down and maintenance of the historically steep shape of the rates curve. The fastest credit spreads tightening coincided with the rates shift rebound in 2009.

However, the co-movement of rates and spreads from 2011 onward shows some change. The widening of spreads coincided with the decline in rates (as usual), but also coincided with the Treasury curve flattening, which is unlike the typical historical pattern. We discuss this in more detail later in the report..

Of course, spreads are influenced by many other factors beside the Treasury curve and the macro-economic outlook encoded therein. However, typically the significant trends of the rate changes do get reflected in spread moves. While the anecdotal evidence presented in this section helps in motivating our research project, it is not sufficiently precise to draw conclusions for the future. To do that, we need a robust statistical estimation of co-movements in treasury rates and credit spreads, which we undertake in the next section.

### 3. ESTIMATES FROM THE MULTI-FACTOR RISK MODEL

To quantify the joint behavior of interest rates and credit spreads, we turn to the Barclays POINT® Global Risk Model (see Lazanas et al. [2011] for a good introduction to these type of models). The current approach employs the DTS (duration times spread) methodology to model credit risk (see Silva [2009]). However, the model allows for different risk configurations. In particular, for this report, we use the following decomposition: six Treasury (key-rate) factors and 27 spread factors, from a combination of nine industries times three rating buckets (AAA/AA, A, and BBB)[3].

The model estimates the covariance matrix of all common driving factors as well as the issuer-specific risk of bonds belonging to each industry/rating sector. We analyze the covariance estimates as of June 2013 and discuss their implications for the relationship between rates and spreads.

The multi-factor risk model has different calibrations available. In this paper we use two standard ones: the first weights all past observations equally, while the second is an exponential-weighted moving average (12month half live) that overweights recent data relative to more distant historical one. The corresponding versions are referred to as long-term and short-term model, respectively.

In order to take into account the issuer-specific risk and incomplete diversification of typical investor's portfolios, we defined a sector portfolio to consist of 20 equally weighted bonds having on average the same maturity and same OAS as the corresponding sector. By construction of the risk model, such portfolio is not exposed to spread twist or OAS dispersion factors. The sector correlations discussed in this paper are the correlations of OAS changes of these hypothetical sector portfolios with the Treasury shift and twist factors.

The results are shown in Tables 2a-2d for the long-term model estimated as of June 2013, in Tables 3a-3d for the short-term model estimated as of June 2013. For comparison, and to highlight the time variability of estimates, we also show the results estimated at the time of the most recent turning rates environment: Tables 4a-4d show the results for the long-term model, and Tables 5a-5d for the short-term model estimated as of December 2003.

The statistical dependence patterns found in these results are discussed in the rest of this section. Their implications for the duration management of credit portfolios are covered in section 4.

---

[3] We use this decomposition to keep our approach consistent with previous versions of this research and to allow the analysis to be done across different levels of spread, here proxied by different ratings. The qualitative results should be similar across approaches.

### 3.1 The Impact of a Treasury Curve Shift

We start by documenting the effect of the Treasury curve shift on credit spreads. The results demonstrate a strong negative correlation between these variables for each credit sector.. Uniform increase in rates is associated with tighter spreads while uniform drop in interest rates leads to wider spreads.

As an example, we find a –33% correlation for A-rated Banking and Brokerage sector in Table 2a, which implies that if all rates rise by a typical amount (an amount equal to 1 standard deviation of the shift factor), the credit spreads in this sector will likely tighten by amount equal to 0.33 of a typical movement (a standard deviation) of the sector's spread factor, all else equal. To translate this statement into nominal levels, we note that the standard deviation of the shift factor, according to the risk model, is 24 bps, and the standard deviation of the A-rated Banking and Brokerage sector spreads is 14 bps, therefore the above prediction is that a 24 bps positive shift in rates will on average translate into almost 5 bps of tightening of the A-rated Banking and Brokerage spreads.

The negative correlation between sector spreads and rates shift is, overall, quite similar across both the short- and long-term versions of the models. However, there are some important differences regarding the range of correlations: they are significantly more disperse on the short-term model, while showing stronger convergence (about -30%) for the longer-term model. Another interesting difference is that the correlations are stable across ratings in the longer-term model while tending to show a negative slope for the short-term model (e.g., correlations are typically more negative for lower rating portfolios).

One can argue that these weaker correlations are due to the effects of the Fed's quantitative easing programme, which has weakened the normal relationships between the economic recovery (represented by the spreads) and monetary policy (represented by the rates). We will see an even stronger evidence of this in the twist factor impact. If this hypothesis is correct, and if one assumes that the QE programme is about to end – taking the economy mechanism closer to their historical norm - then it would follow that the more appropriate numbers for forward looking prediction might be closer to long-term model, or perhaps even the models estimated in 2003 rather than the short-term model of 2013 vintage.

In the end, the particular patterns of dependence of the strength of negative correlation on the sector or credit quality are driven by several factors, including the underlying economics of the corresponding sectors, fiscal and monetary policy, and the varying composition of the Credit Index, which occasionally has a greater representation of certain types of companies in a particular rating class. These shifts were particularly visible after the financial crisis. Many companies were downgraded from AAA/AA to A or even the BBB category, thus changing the compositions of those baskets and their dependence.

Some of the dependence patterns in correlations between industry sector/rating category spreads and interest rates shift factor, which remain valid across time and model flavors, are:

- The cyclical industries exhibit stronger negative correlation with the shift factor than do non-cyclical industries. This should come as no big surprise, as by definition the dependence of the cyclical industries on economic decline or recovery, reflected by the changing levels of interest rates, is indeed stronger.

- In most industries, with the exception of Banking and Brokerage and Consumer sectors, lower credit quality is associated with greater degree of negative correlation. This is rather intuitive, because the companies with lower credit quality are typically more affected by the changes in the economic outlook, as reflected in the general level of interest rates.

Equally telling are some of the dependence patterns in correlations which substantially changed with time and depend strongly on model variants:

- In the years prior to the financial crisis, the Financials sector used to uniformly exhibit a pattern of the higher credit ratings being associated with greater degree of negative correlation (see Tables 4a and 5a). After the crisis, the pattern changed -- the correlations in the Banking and Brokerage sector are now almost independent of the rating level, and those in the Financial Companies, Insurance and REITs are actually strongly increasing with the lower rating (see Tables 2a and 3a), closer to the pattern seen for other industries.

- Before the crisis, the short- and long-term models showed a similar variability of correlations across sectors and ratings. After the crisis, markets tended to move more in tandem, and long-term variability decreased. Only recently (short-term model) do we see an increased range of behavior, more consistent with historical patterns.

### 3.2 The Impact of a Treasury Curve Twist

We now discuss the effect of the Treasury curve twist on credit spreads. One of the biggest casualties of the financial crisis and subsequent QE-filled years was the statistical dependence of credit sectors on the Treasury twist factor.

We used to see a strong positive correlation of credit spreads across all sectors and ratings with the steepening yield curve (see Tables 4b and 5b). That is, a steepening of the curve is associated with higher spreads. This was consistent with what we observed during the periods at the bottom of the economic cycle where the curve inversions and steepening were driven by the Fed actions at the short end of the curve.

The moderate flattening of the curve caused by the start of the QE a few years ago coincided with some spread widening with uncertainty about the strength of the recovery, leading to negative correlations between Treasury twists and spread, which was contrary to the normal pattern. More recently, the further steepening of the curve due to concerns about the end of QE – while the short end is still nailed down by the Fed's near-zero interest rate policy – coincided with a modest credit pickup, again in contrast to the long-term norm. This is most obvious if one contrasts the results from the more recent short-term model in Table 3b with those from the short-term model as of 2003, in Table 5b. The effect is also seen in long-term calibrations, but to a lesser extent. In this regard, changes in monetary policy can have a significant effect on how portfolios react to changes in interest rates.

### 3.3 The Treasury Curve Impact On Non-Sector Credit Factors

The patterns of the impact of the Treasury curve shift and twist on credit curve shape and dispersion factors have also significantly changed in the recent years.

We used to observe a positive correlation of the credit curve slope with rates shift, i.e. the rising rates were generally associated with steepening credit spreads. This seemed natural and in line with overall logic of the rising rates being associated with improving fundamentals -- not only those usually lead to tighter spread curves but also to more upward sloping ones. These same better fundamentals would normally lead to tighter dispersion of the spreads as well (a negative correlation of the credit dispersion factor), as seen in Tables 4c and 5c.

Similarly, we used to see a negative correlation of the credit spread slope, and a positive correlation of the spread dispersion with the Treasury twist factor, with the explanation being roughly the same.

However, the more recent QE-influenced period has resulted in all of the correlations being negative, as seen in Table 3c. The recent estimate of the long-term model is also partially influenced by these past several years of the unusual behavior (see Table 2c).

So the main unchanged pattern is that the credit spread dispersion is still negatively correlated with interest rate levels, which makes sense -- the rising rates still mean economic recovery and it still results in relative convergence of issuer spreads and smaller risk premia for issuer specific uncertainty.

The negative correlation of the spread curve slope and Treasury curve twist is also intact, despite the meaning of the latter being somewhat subverted in the past few years. Still, a steepening yield curve is not so good for longer maturity credit risk premia -- whether it happens due to Fed lowering the short rates or the Fed no longer being able to hold still the long rates due to impending end of QE and fears of inflation.

### DURATION MANAGEMENT OF CREDIT PORTFOLIOS

The results of our study have important implications for risk management as well as for identifying relative value opportunities across sectors with different interest-rate sensitivities.

The directionality of credit spreads and interest rates poses a challenge to credit investors who want to manage the interest rate exposure of their portfolio. Because spreads tend to move in conjunction with underlying interest rates, a corporate bond is not fully insulated from rate movements if hedged with the same-duration Treasury bond. In other words, a credit bond portfolio benchmarked against government bond index (such as the overweight credit portion of a typical fixed income portfolio) will not be neutral to interest rate movement if it has a matching duration with the Treasury benchmark.

Indeed, duration measures the sensitivity of bond price with respect to the change in yield. For a given shift in interest rates, the corresponding change in the corporate yield is smaller because it gets partially offset by the tightening of the spread. To account for this fact, we introduce the concept of *Effective Duration*. We define it as the sensitivity of corporate bond prices to changes in the interest rate component of the yield.

The multi-factor risk model allows us to estimate the volatility $\sigma_F$ of the shift and twist factor of the yield curve as well as the volatility $\sigma_S$ of the typical industry/rating sector portfolio spread and its correlation $\rho(S, F_{treas})$ with the rate factors. Given these values, the expected change in spread given the change in the treasury factor (either shift or twist) is:

$$\frac{\Delta S}{\sigma_S} = \rho(S, F_{treas}) \cdot \frac{\Delta F_{treas}}{\sigma_F}$$

Using this relationship, we can estimate the price impact of the parallel shift on the credit bond using the chain rule:

$$\frac{1}{P} \cdot \frac{\Delta P}{\Delta F_{shift}} = \frac{1}{P} \cdot \frac{\Delta P}{\Delta Y} \cdot \frac{\Delta Y}{\Delta F_{shift}} + \frac{1}{P} \cdot \frac{\Delta P}{\Delta S} \cdot \frac{\Delta S}{\Delta F_{shift}}$$

In the first term in the left hand side we introduced the change in the underlying yield of the Treasury curve, which by construction is assumed to be same as the change in the shift factor when the parallel shift is the sole movement of the yield curve. Therefore $\Delta Y / \Delta F_{shift} = 1$.

The fractional change in price with respect to change in yield is, by definition, the modified duration of the bond (with negative sign).

In the second term, we introduced the spread, whose relationship with the shift factor we explained above. The fractional change in price with respect to change in spread is, by definition, the spread duration of the bond (with negative sign).

Defining the fractional change in price with respect to change in shift factor as the effective duration (with negative sign) we obtain:

$$D_{eff} = D_{mod} + \rho\left(S, F_{shift}\right) \cdot \frac{\sigma_{spread}}{\sigma_{shift}} \cdot D_{spread}$$

Here, $D_{eff}$ stands for the effective duration, $D_{mod}$ is the modified duration, $D_{spread}$ is the spread duration, $\rho$ is the correlation between spreads and Treasury shift, $\sigma_{spread}$ is the volatility of spreads, and $\sigma_{shift}$ is the volatility of the Treasury shift factor (both volatilities must be measured in absolute terms and expressed in equal units, e.g. bp/month).

Since the correlation of spreads and yields is negative and quite substantial, the effective duration will be typically smaller than modified duration. For most fixed coupon bonds modified duration and spread duration differ very slightly, hence the effective duration is approximately equal to a fraction of the modified duration. We denote this fraction as *Effective Duration Multiplier* $M_{eff}$, and rewrite the effective duration definition as follows:

$$D_{eff} \approx M_{eff} \cdot D_{mod}$$

$$M_{eff} = 1 + \rho\left(S, F_{shift}\right) \cdot \frac{\sigma_{spread}}{\sigma_{shift}}$$

The estimated values of the effective duration multiplier are shown in Tables 2d, 3d, 4d and 5d, for each of the estimates of the risk model, respectively. To illustrate with an example, let's look at the results for the long-term risk model of 2013 vintage (Table 2d), and consider two 10-year par bonds – a Treasury and a typical corporate bond in A-rated Consumer Cyclicals. Suppose both have modified duration of 7.5 years, the spread duration of the corporate bond is also 7.5 years.

We observe that the correlation between the 10-year yield and the spread on the corporate is -34%. This means that a 10 bps increase in Treasury rates will be typically accompanied by a decrease in the spread of the corporate bond, equal to the correlation multiplied by the ratios of the standard deviations of spreads and rates factors. The standard deviation of the rate shifts is 24.3 bp/month (as determined from the Barclays POINT® risk model), and the standard deviation of the spreads in A Consumer Cyclicals is 18.2 bp/month. Therefore, the corresponding spread tightening, predicted by the risk model, is equal to 10 bps * 34% * 18.2 / 24.3 = 2.5 bps.

The price impact of the 10 bp increase in rates is 7.5 * 0.10 = 0.75 decrease in price per 100 initial value in both bonds. However, in case of the corporate bond this price decrease will be offset by 2.5 bp decrease in spreads, and associated price impact of 7.5 * 0.025 = 0.1875 per 100 initial value. Thus the price of the corporate bond will decrease only by 0.75 – 0.1875 = 0.5625. Since this price change was effected by a 10 bp rise in rates, the effective duration is 0.5625 / 0.10 = 5.625 years. This effective duration value represents 75% of the original modified duration of 7.5 years (as reported in the figure).

Thus, a credit portfolio that is overweight in this corporate bond, while benchmarked to a Treasury portfolio with matching modified duration will in fact be mismatched in terms of effective duration, and consequently in terms of expected sensitivity to interest rate moves.

Another interesting take-away from this analysis is related to the banking and brokerage portfolios. The effective duration of these portfolios is significantly lower in 2013 (compared with 2003), especially in the short-term model. As discuss previously, this may be the consequence of the atypical behaviour this industry has registered since the financial crisis.

We would like to emphasize that when measuring the risk of credit portfolios within the Barclays POINT portfolio analytics system the effect of the correlation between the credit spreads and Treasury rates is fully taken into account by virtue of using the complete multi-factor risk model with full covariance matrix of dependencies. The exposition above illustrates the source of high contribution of interest rate risks to the tracking error of many credit portfolios even when they are apparently well balanced in terms of modified duration.

Many credit portfolio managers are not actively managing the duration or curve position of their portfolios, but are instead following the constraints imposed by broader multi-asset class and duration allocations within risk budgeting frameworks of aggregate fixed income portfolios. In such cases either the portfolio managers responsible for asset allocation can take into account the rates-spreads directionality in setting the goals for the credit PMs, or the credit portfolio managers can explicitly adjust their duration targets if the implicit assumption in the asset allocation process is that of independence of rates and spreads.

## CONCLUSIONS

In this study we used the statistically robust framework of the Barclays POINT® Global Risk Model to analyze the co-movements of interest rates and credit spreads. The main message is that both shifts and twists of the Treasury yield curve are accompanied by significant changes in both the level and slope of the credit spread curve.

We would like to reiterate that our study concerns contemporaneous correlations and is not, by itself, a statement of causal relationship. Rather, the existence and robustness of correlations across a long historical period from 1990 until the present can be taken as a evidence for the common economic driving factors between rates and spreads.

Portfolio managers need to consider the rates-spreads directionality effects when fine-tuning their interest-rate hedging strategies and relative value decisions across credit sectors in the environment when credit specific news are dominated by macro-economic news leading to significant Treasury curve moves.

The intervening years since our original studies saw periods ranging from very low risk (2005-2006) to extremely high risk (2008) and subsequent recovery accompanied by the peculiar experience of the Fed's quantitative easing influencing both interest rates and credit markets. As discussed in section 3, some of the results (such as the negative correlation of spreads and Treasury curve shifts) remain quite robust, while others (the correlation of the spreads with Treasury curve twists) become dislocated or even change the sign.

Although we do not give any specific forecasts in this paper, we would caution the readers to carefully choose their scenarios and pick those results which they think will be better representative of the near future, while applying this framework for credit portfolio management. Whether the most recent estimates will continue to hold depends on the assumption that the economic conditions and the Fed's actions influencing the shape of the Treasury curve will continue to remain the same. If one thinks that these conditions will change, then it is entirely possible that the more representative statistics for the future may be found in the more distant past, rather than recently.

**Table 2a: Industry Portfolio Spread Correlations with Treasury Curve Shifts (Long-Term Model)**

|  | AAA/AA | A | BBB |
|---|---|---|---|
| **FINANCIALS** |  |  |  |
| Banking and Brokerage | -32% | -33% | -31% |
| Financial Companies, Insurance and REITS | -26% | -33% | -38% |
|  |  |  |  |
| **INDUSTRIALS** |  |  |  |
| Basic Industries and Capital Goods | -32% | -35% | -35% |
| Consumer Cyclicals | -38% | -34% | -30% |
| Consumer Non-Cyclicals | -35% | -32% | -30% |
| Communication and Technology | -31% | -34% | -36% |
| Energy and Transportation | -37% | -37% | -38% |
|  |  |  |  |
| UTILITIES | -24% | -35% | -34% |
| NON-CORPORATE | -32% | -34% | -36% |

**Table 2b: Industry Portfolio Spread Correlations with Treasury Curve Twists (Long-Term Model)**

|  | AAA/AA | A | BBB |
|---|---|---|---|
| **FINANCIALS** |  |  |  |
| Banking and Brokerage | 13% | 13% | 13% |
| Financial Companies, Insurance and REITS | 11% | 13% | 12% |
|  |  |  |  |
| **INDUSTRIALS** |  |  |  |
| Basic Industries and Capital Goods | 11% | 12% | 13% |
| Consumer Cyclicals | 13% | 13% | 14% |
| Consumer Non-Cyclicals | 12% | 12% | 12% |
| Communication and Technology | 9% | 13% | 14% |
| Energy and Transportation | 12% | 13% | 14% |
|  |  |  |  |
| UTILITIES | 10% | 12% | 13% |
| NON-CORPORATE | 8% | 13% | 14% |

**Table 2c: Additional Spread Factor Correlations with Treasury Curve Changes (Long-Term Model)**

|  | Treasury Shift | Treasury Twist |
|---|---|---|
| Credit Spread Twist (Steepening) | 5% | -1% |
| Credit Spread Dispersion | -42% | 17% |

**Table 2d: Effective Duration Multipliers for Industry / Rating Sectors (Long-Term Model)**

|  | AAA/AA | A | BBB |
|---|---|---|---|
| **FINANCIALS** |  |  |  |
| Banking and Brokerage | 79% | 81% | 65% |
| Financial Companies, Insurance and REITS | 83% | 69% | 46% |
|  |  |  |  |
| **INDUSTRIALS** |  |  |  |
| Basic Industries and Capital Goods | 87% | 79% | 67% |
| Consumer Cyclicals | 84% | 75% | 63% |
| Consumer Non-Cyclicals | 84% | 82% | 77% |
| Communication and Technology | 88% | 74% | 59% |
| Energy and Transportation | 82% | 79% | 70% |
|  |  |  |  |
| UTILITIES | 87% | 79% | 69% |
| NON-CORPORATE | 91% | 82% | 66% |

**Table 3a: Industry Portfolio Spread Correlations with Treasury Curve Shifts (Short-Term Model)**

|  | AAA/AA | A | BBB |
|---|---|---|---|
| **FINANCIALS** |  |  |  |
| Banking and Brokerage | -39% | -34% | -38% |
| Financial Companies, Insurance and REITS | -21% | -34% | -42% |
|  |  |  |  |
| **INDUSTRIALS** |  |  |  |
| Basic Industries and Capital Goods | -25% | -26% | -36% |
| Consumer Cyclicals | -29% | -27% | -32% |
| Consumer Non-Cyclicals | -25% | -23% | -26% |
| Communication and Technology | -19% | -29% | -37% |
| Energy and Transportation | -21% | -31% | -36% |
|  |  |  |  |
| **UTILITIES** | -34% | -29% | -30% |
| **NON-CORPORATE** | -23% | -36% | -15% |

**Table 3b: Industry Portfolio Spread Correlations with Treasury Curve Twists (Short-Term Model)**

|  | AAA/AA | A | BBB |
|---|---|---|---|
| **FINANCIALS** |  |  |  |
| Banking and Brokerage | -26% | -24% | -26% |
| Financial Companies, Insurance and REITS | -16% | -23% | -28% |
|  |  |  |  |
| **INDUSTRIALS** |  |  |  |
| Basic Industries and Capital Goods | -18% | -18% | -24% |
| Consumer Cyclicals | -20% | -19% | -22% |
| Consumer Non-Cyclicals | -18% | -17% | -19% |
| Communication and Technology | -14% | -20% | -25% |
| Energy and Transportation | -15% | -22% | -25% |
|  |  |  |  |
| **UTILITIES** | -23% | -20% | -21% |
| **NON-CORPORATE** | -14% | -25% | -12% |

**Table 3c: Additional Spread Factor Correlations with Treasury Curve Changes (Short-Term Model)**

|  | Treasury Shift | Treasury Twist |
|---|---|---|
| Credit Spread Twist (Steepening) | -21% | -13% |
| Credit Spread Dispersion | -50% | -33% |

**Table 3d: Effective Duration Multipliers for Industry / Rating Sectors (Short-Term Model)**

|  | AAA/AA | A | BBB |
|---|---|---|---|
| **FINANCIALS** |  |  |  |
| Banking and Brokerage | 68% | 70% | 48% |
| Financial Companies, Insurance and REITS | 83% | 65% | 32% |
|  |  |  |  |
| **INDUSTRIALS** |  |  |  |
| Basic Industries and Capital Goods | 88% | 84% | 62% |
| Consumer Cyclicals | 85% | 79% | 63% |
| Consumer Non-Cyclicals | 88% | 86% | 78% |
| Communication and Technology | 92% | 75% | 55% |
| Energy and Transportation | 88% | 80% | 66% |
|  |  |  |  |
| **UTILITIES** | 79% | 83% | 73% |
| **NON-CORPORATE** | 93% | 75% | 82% |

**Table 4a: Industry Portfolio Spread Correlations with Treasury Curve Shifts (Long-Term Model, 2003)**

|  | AAA/AA | A | BBB |
|---|---|---|---|
| FINANCIALS |  |  |  |
| Banking and Brokerage | -31% | -31% | -22% |
| Financial Companies, Insurance and REITS | -38% | -31% | -29% |
|  |  |  |  |
| INDUSTRIALS |  |  |  |
| Basic Industries and Capital Goods | -31% | -43% | -36% |
| Consumer Cyclicals | -41% | -41% | -22% |
| Consumer Non-Cyclicals | -40% | -35% | -33% |
| Communication and Technology | -31% | -38% | -31% |
| Energy and Transportation | -41% | -43% | -40% |
|  |  |  |  |
| UTILITIES | -21% | -36% | -29% |
| NON-CORPORATE | -31% | -35% | -41% |

**Table 4b: Industry Portfolio Spread Correlations with Treasury Curve Twists (Long-Term Model, 2003)**

|  | AAA/AA | A | BBB |
|---|---|---|---|
| FINANCIALS |  |  |  |
| Banking and Brokerage | 17% | 19% | 15% |
| Financial Companies, Insurance and REITS | 20% | 18% | 16% |
|  |  |  |  |
| INDUSTRIALS |  |  |  |
| Basic Industries and Capital Goods | 15% | 21% | 20% |
| Consumer Cyclicals | 19% | 21% | 15% |
| Consumer Non-Cyclicals | 19% | 17% | 18% |
| Communication and Technology | 16% | 20% | 18% |
| Energy and Transportation | 18% | 21% | 21% |
|  |  |  |  |
| UTILITIES | 11% | 18% | 16% |
| NON-CORPORATE | 15% | 18% | 21% |

**Table 4c: Additional Spread Factor Correlations with Treasury Curve Changes (Long-Term Model, 20**

|  | Treasury Shift | Treasury Twist |
|---|---|---|
| Credit Spread Twist (Steepening) | 10% | -13% |
| Credit Spread Dispersion | -56% | 48% |

**Table 4d: Effective Duration Multipliers for Industry / Rating Sectors (Long-Term Model, 2004)**

|  | AAA/AA | A | BBB |
|---|---|---|---|
| FINANCIALS |  |  |  |
| Banking and Brokerage | 89% | 87% | 81% |
| Financial Companies, Insurance and REITS | 88% | 87% | 84% |
|  |  |  |  |
| INDUSTRIALS |  |  |  |
| Basic Industries and Capital Goods | 92% | 87% | 84% |
| Consumer Cyclicals | 89% | 83% | 79% |
| Consumer Non-Cyclicals | 89% | 89% | 87% |
| Communication and Technology | 89% | 84% | 78% |
| Energy and Transportation | 87% | 86% | 82% |
|  |  |  |  |
| UTILITIES | 93% | 87% | 81% |
| NON-CORPORATE | 93% | 88% | 71% |

**Table 5a: Industry Portfolio Spread Correlations with Treasury Curve Shifts (Short-Term Model, 2003)**

|  | AAA/AA | A | BBB |
|---|---|---|---|
| **FINANCIALS** | | | |
| Banking and Brokerage | -56% | -52% | -46% |
| Financial Companies, Insurance and REITS | -52% | -43% | -45% |
| | | | |
| **INDUSTRIALS** | | | |
| Basic Industries and Capital Goods | -61% | -62% | -56% |
| Consumer Cyclicals | -58% | -56% | -49% |
| Consumer Non-Cyclicals | -57% | -53% | -55% |
| Communication and Technology | -44% | -51% | -45% |
| Energy and Transportation | -57% | -60% | -60% |
| | | | |
| UTILITIES | -56% | -54% | -40% |
| NON-CORPORATE | -60% | -53% | -55% |

**Table 5b: Industry Portfolio Spread Correlations with Treasury Curve Twists (Short-Term Model, 2003**

|  | AAA/AA | A | BBB |
|---|---|---|---|
| **FINANCIALS** | | | |
| Banking and Brokerage | 33% | 31% | 30% |
| Financial Companies, Insurance and REITS | 35% | 33% | 23% |
| | | | |
| **INDUSTRIALS** | | | |
| Basic Industries and Capital Goods | 23% | 28% | 31% |
| Consumer Cyclicals | 22% | 32% | 40% |
| Consumer Non-Cyclicals | 25% | 23% | 25% |
| Communication and Technology | 26% | 32% | 36% |
| Energy and Transportation | 26% | 29% | 30% |
| | | | |
| UTILITIES | 27% | 31% | 33% |
| NON-CORPORATE | 21% | 26% | 36% |

**Table 5c: Additional Spread Factor Correlations with Treasury Curve Changes (Short-Term Model, 20**

|  | Treasury Shift | Treasury Twist |
|---|---|---|
| Credit Spread Twist (Steepening) | 9% | -12% |
| Credit Spread Dispersion | -46% | 44% |

**Table 5d: Effective Duration Multipliers for Industry / Rating Sectors (Short-Term Model, 2003)**

|  | AAA/AA | A | BBB |
|---|---|---|---|
| **FINANCIALS** | | | |
| Banking and Brokerage | 84% | 83% | 79% |
| Financial Companies, Insurance and REITS | 85% | 79% | 75% |
| | | | |
| **INDUSTRIALS** | | | |
| Basic Industries and Capital Goods | 83% | 79% | 75% |
| Consumer Cyclicals | 85% | 75% | 64% |
| Consumer Non-Cyclicals | 83% | 83% | 79% |
| Communication and Technology | 80% | 76% | 58% |
| Energy and Transportation | 80% | 80% | 77% |
| | | | |
| UTILITIES | 77% | 77% | 62% |
| NON-CORPORATE | 89% | 86% | 61% |